\documentclass[a4paper]{report}
\usepackage[utf8]{inputenc}
\usepackage[T1]{fontenc}
\usepackage{RJournal}
\usepackage{amsmath,amssymb,array}
\usepackage{booktabs}

\begin{document}

%% do not edit, for illustration only
\sectionhead{Contributed research article}
\volume{XX}
\volnumber{YY}
\year{20ZZ}
\month{AAAA}

%% replace RJtemplate with your article
\begin{article}
  % !TeX root = RJwrapper.tex
\title{iotools: High-Performance I/O Tools for \R}
\author{by Taylor Arnold, Michael J. Kane, and Simon Urbanek}

\maketitle

\abstract{
The iotools package provides Input/Output (I/O) intensive
data processing in \R~\citep{R}.  Efficient parsing methods
are included to minimize copying and avoid the use of
intermediate strings. Methods support ``chunk-wise'' 
operations for computing on streaming input including arbitrarily large
files.  We present a set of examples for \pkg{iotools}, as well as 
benchmarks comparing similar functions provided in both core-\R~as well as
other contributed packages.
}

\section{Introduction}

When processing large data sets on a single machine 
the performance bottleneck is often getting data from the hard drive to the 
format required by the programming environment. The associated latency 
comes from a combination of two sources. First, there is hardware
latency from moving data from the hard-drive to RAM. This is especially
the case with ``spinning'' disk drives, which can have throughput speeds
several orders of magnitude less than those of RAM. 
Hardware approaches for addressing latency have been an active area of 
research and development since hard-drives have existed.
Solid state drives and redundant arrays of inexpensive disks (RAID) now 
provide throughput comparable to RAM; they are readily
available on commodity systems; and they continue to improve.  
The second source comes from 
the software latency associated with transforming data from its 
representation
on the disk to the format required by the programming 
environment.  This translation drags performance for many \R~
users, especially in the context of larger data sources. 

The code below uses the \pkg{microbenchmark} \citep{microbenchmark} package
to compare the time needed to read, parse, and create a
\code{data.frame} with the time needed to simply read data from disk.
The file contains comma-separated 
value (CSV) file with 29 columns and 7,009,728 rows.
It takes about 20 times longer to perform the former compared to the
latter indicating there may be room for improvement.

\begin{example}
> library(microbenchmark)
> col_classes = c(rep("integer", 8), "character", "integer", "character",
+                 rep("integer", 5), "character", "character", 
+                 rep("integer", 4), "character", rep("integer", 6))
>
> f = "2008.csv"
> microbenchmark(s=read.csv(f, colClasses=col_classes), unit="s", times=5)
Unit: seconds
 expr     min       lq     mean   median       uq      max neval
    s 91.7096 92.08713 92.43579 92.24694 92.92127 93.21401     5
> microbenchmark(s=readBin(f, "raw", file.info(f)$size), times=5, unit="s")
Unit: seconds
 expr       min        lq      mean    median        uq       max neval
    s 0.4596488 0.4817199 0.4857666 0.4940194 0.4940351 0.4994098     5
\end{example}

This is not to say \code{read.csv} and its associated functions
are poorly written. On the contrary, they are robust and do an excellent job 
inferring data format and shape characteristics. They 
allow users to import and examine a data set without
knowing how many rows it has, how many columns it has, or its column types.
Because of these function statisticians using \R~are able to focus on data 
exploration and modeling instead of file formats and schemas.

While these functions are sufficient for processing relatively small
data sets, larger ones require a different approach.
For large files, data are often processed on a single 
machine by extracting consecutive rows or ``chunks'' from the filesystem, 
a chunk is processed, and then the next chunk is retrieved.
The results from processing each chunk 
are then aggregated and returned. Small, manageable subsets are 
streamed from the disk to the processor and only require the memory
needed to represent a single chunk.
%This approach has become increasingly performant on commodity hardware due
%to decreased hard-drive latency and increased bandwidth.
%in the last five to 10
%years with solid state drives providing throughput comparable to RAM
%and even further back with redundant arrays of inexpensive disks providing
%multiple channels from the disk to RAM.

This approach is common not only on single machine but also 
in distributed environements with technologies like Spark \citep{Spark}
and Hadoop MapReduce \citep{Dean2008}. Clusters of commodity machines are 
able to process
vast amounts of data one chunk at a time. Statistical methodology is
compatible with this computational approach and is justified 
in a variety of 
statistical/machine learning contexts including \cite{Hartigan1975},
\cite{Kleiner2011}, \cite{Guha2012}, and \cite{Matloff2014} to name a few.

However, \R~currently does not address this common computing pattern.
Packages such as \pkg{bigmemory} \citep{Kane2013} and \pkg{ff}
\citep{ff} provide data structures using their own binary format on a disk.
They make use of memory-mapped files that may be stored on disk. 
The data structures they provide are not native \R~objects. 
They do not exhibit copy-on-write behavior. And, in general, they cannot be 
seamlessly integrated with \R's plethora of user contributed packages.
The \pkg{readr} package \citep{readr} provides fast importing of 
\code{data.frame} objects but it does not support chunk-wise operations
for arbitrarily large files.
The \pkg{foreach} package \citep{foreach}, and it's associated 
\pkg{iterators} package \citep{iterators}, 
provide a general framework for 
chunked processing but does not provide the low-level connection-based 
utilities 
for transforming binary data stored on a disk to those native to \R. 

The \pkg{iotools} package provides in-place stream processing for
any data source represented as a connection.
Users of the package can 
import text and binary data into \R~and process large data sets as
chunks. The package can be several orders of magnitude faster
when compared to \R's native facilities. The package provides general
tools for quickly processing large data sets in consecutive chunks, both in- 
and out-of-core, and provides a basis for speeding up distributed computing 
frameworks including Hadoop Streaming \citep{Hadoop} and Spark.

The rest of this paper introduces the use of the \pkg{iotools} package 
for quickly importing data from disk to \R~and processing those data.
Examples center
around calculation of OLS slope coefficients via the normal
equations. This particular calculation was chosen because it balances
read/write times with processing time. 

\section{A note on the data used in this paper}

Examples in this paper make use of the ``Airline on-time performance'' data
set \citep{AirlineDataSet}, which was released for the 2009
American Statistical Association (ASA) Section on Statistical
Computing and Statistical Graphics biannual data exposition.
The data set includes commercial flight arrival and departure information
from October 1987 to April 2008 for those carriers with at least 1\% of
domestic U.S. flights in a given year. In total, there is
information for over 120 million flights, with 29 variables related to flight
time, delay time, departure airport, arrival airport, and so on.
In total, the uncompressed data set is 12 gigabytes (GB) in size.

It should be noted the $12\,$GB Airline On-time data set will likely not be
considered ``big'' to many readers.  However, in designing the
examples two principles were considered before sheer data size.
First, the data set is publicly available. The code included in the
Supplemental Material of this paper is capable of downloading the
data set and running the benchmarks. Users are encouraged to engage
the data themselves by trying the code examples and developing their 
own analyses. Second, the
data set is large enough to investigate the performance 
properties of \pkg{iotools} along with it's associated scaling behavior.
Together, the data set and the code available with this paper
provide a set of accessible and reproducible examples forming
a basis for instruction and subsequent development.

\section{I/O Methods and Formatters}

\R's file operations make use of Standard C 
input/output operations includig \code{fread} and \code{fwrite}.
Data are read in, elements are parsed, and parsed values populate
data structures.
The \pkg{iotools} package also uses the Standard C library
but it makes use of ``bulk'' binary operations including \code{memchr}
and \code{strchr}. These functions make use of hardware specific, single
instruction, multiple data operations (SIMD) and tend to be faster than 
their Standard I/O counterpart, which uses \code{fread} with search
functions in the user-space. As a result \pkg{iotools} is able to find and
retrieve data
at a higher rate. In addition, an entire data set or chunk is buffered rather
than scanned and transformed line-by-line as in the \code{read.table} function.
Thus, by buffering chunks of data and 
making use of low-level, system functions \pkg{iotools} is able to 
provide more performant data ingestion than what is available in 
base \R as well as other packages.

%\subsection{\code{dstrsplit}}
\subsection{Importing data with \code{read.csv.raw} and \code{dstrsplit}}

In this section the \pkg{iotools} import functionality is applied 
to the airline data files, each of 
which is csv-formatted. Files begin with header
and column types consistent across each of the 22 files. Each file
corresponds to a full year of data, except the first year (1987), where
the data starts on October 14th. Importing 1987 flights with \pkg{iotools}
is shown below. The \code{readAsRaw} function takes either connection or
file name and returns the contents as a \code{raw} type. The \code{dstrsplit}
function parses the a raw vector according to the specified column types
and returns a \code{data.frame}. Since these functions may be considered
``lower-level,'' the \code{read.csv.raw} was written for importing data
in a manner similar to \code{read.table}. It supports similar parameters
but runs faster.

\begin{example} 
> # read.table with column types specified
> microbenchmark(read.table("1987.csv", header=TRUE, sep=",", 
+                           colClasses=col_classes), times=5, unit="s")
Unit: seconds
 expr      min       lq     mean   median       uq      max neval
    s 12.85797 12.93201 13.50017 13.10885 13.91895 14.68304     5
>
> # iotools with column types specified
> microbenchmark(dstrsplit(readAsRaw(file.path(path, "1987.csv")), sep=",",
+                          col_types=col_classes), times=5, unit="s")
+                times=5, unit="s")
Unit: seconds
 expr      min       lq     mean   median       uq      max neval
    s 2.671205 2.728222 2.733358 2.740685 2.761237 2.765441     5
> 
> # read.table with column types inferred
> microbenchmark(read.table("1987.csv", header=TRUE, sep=","),times=5, unit="s")
Unit: seconds
 expr      min       lq     mean   median       uq      max neval
    s 15.48836 15.58127 15.59946 15.60643 15.63559 15.68568     5
>
> # iotools with column types inferred 
> microbenchmark(read.csv.raw("1987.csv", header=TRUE, sep=","),
+                             times=5, unit="s")
Unit: seconds
 expr      min       lq     mean   median       uq      max neval
    s 2.705287 2.781435 2.798451 2.806583 2.832832 2.866119     5
\end{example}

\begin{figure}[htbp]
\centering
\includegraphics[width=5in]{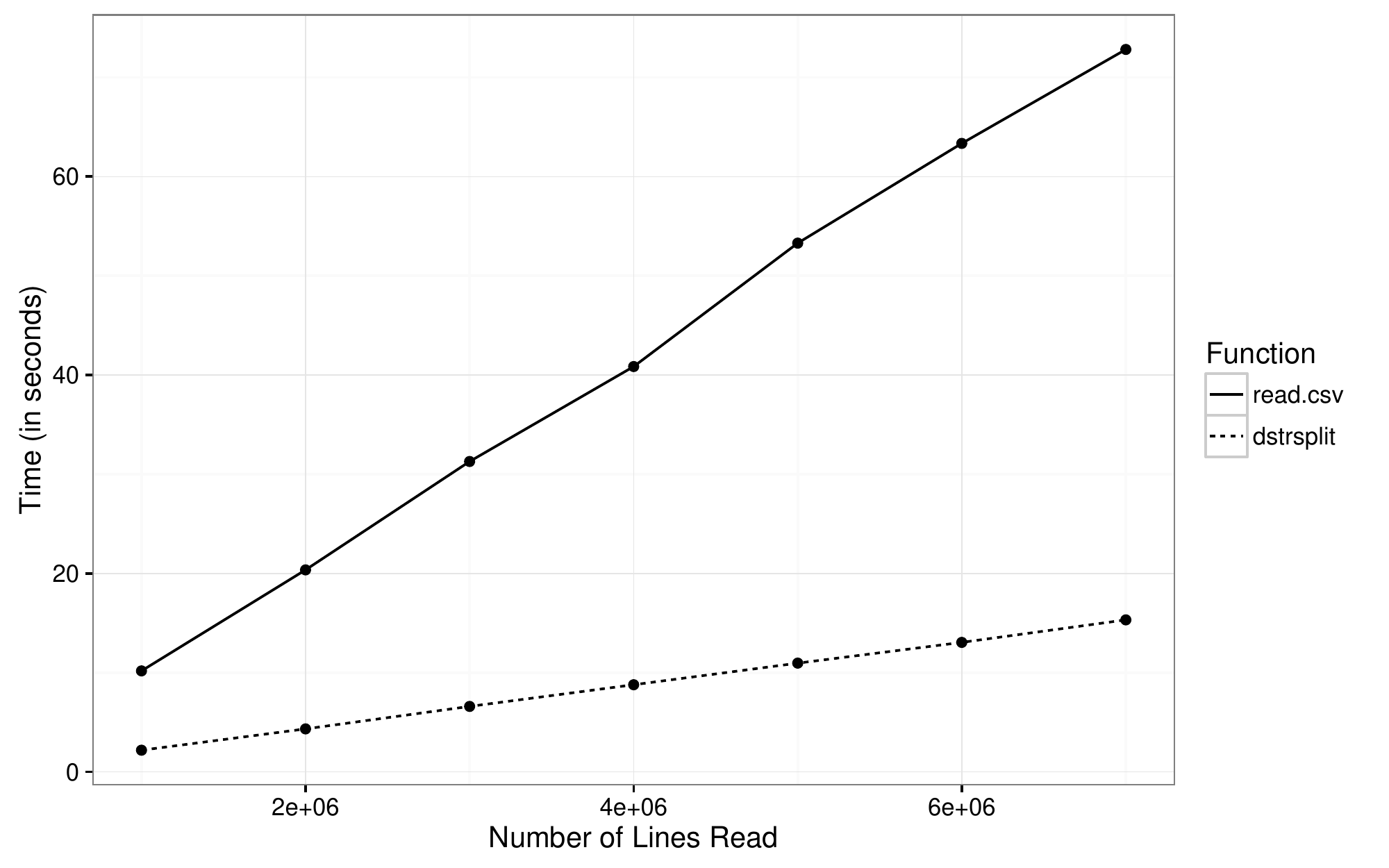}
\caption{Timings using \code{read.csv}, and \code{dstrsplit}}
\label{fig:bench_dstrsplit}
\end{figure}

The performance of \code{read.csv} and \code{dstrsplit} 
is compared in Figure \ref{fig:bench_dstrsplit}. 
The benchmark measures the import times for 1,000,000 to 7,000,000
lines\footnote{Benchmarks were performed on a MacBook
Pro with 2.7GHz Intel Core i7 processor (4 physical cores, 8 logical cores) 
with 16 GB 1600 MHz DDR3 RAM and
Flash Storage.}. The visualization shows importing data using
\code{read.csv} takes about five times longer than \code{dstrsplit}.

The \code{dstrsplit} function takes either
a \code{raw} or \code{character} vector and splits it 
into a data frame according to a specified separator. The columns
may be of type \code{logical}, \code{integer}, \code{numeric}, 
\code{character}, \code{raw}, \code{complex}, 
\code{POSIXct}, and \code{NA} where \code{NA} indicates the column
should be skipped in the output.  It may be considered a building 
block for both \code{read.csv.raw} as well as other computing infrastructures
including Hadoop, pipes, and database connections to name a few.

It should be noted \code{factor} types are not supported. It will
be shown later \code{dstrsplit} can be used in a streaming context
and in this case data are read sequentially. As a result, the 
set of factor levels cannot be deduced until the entire sequence 
is read. However, in most cases, a caller knows the schema 
and is willing to specify factor levels or the caller is willing to 
use a single pass through the data to find factor levels. 

\begin{figure}[htbp]
\centering
\includegraphics[width=5in]{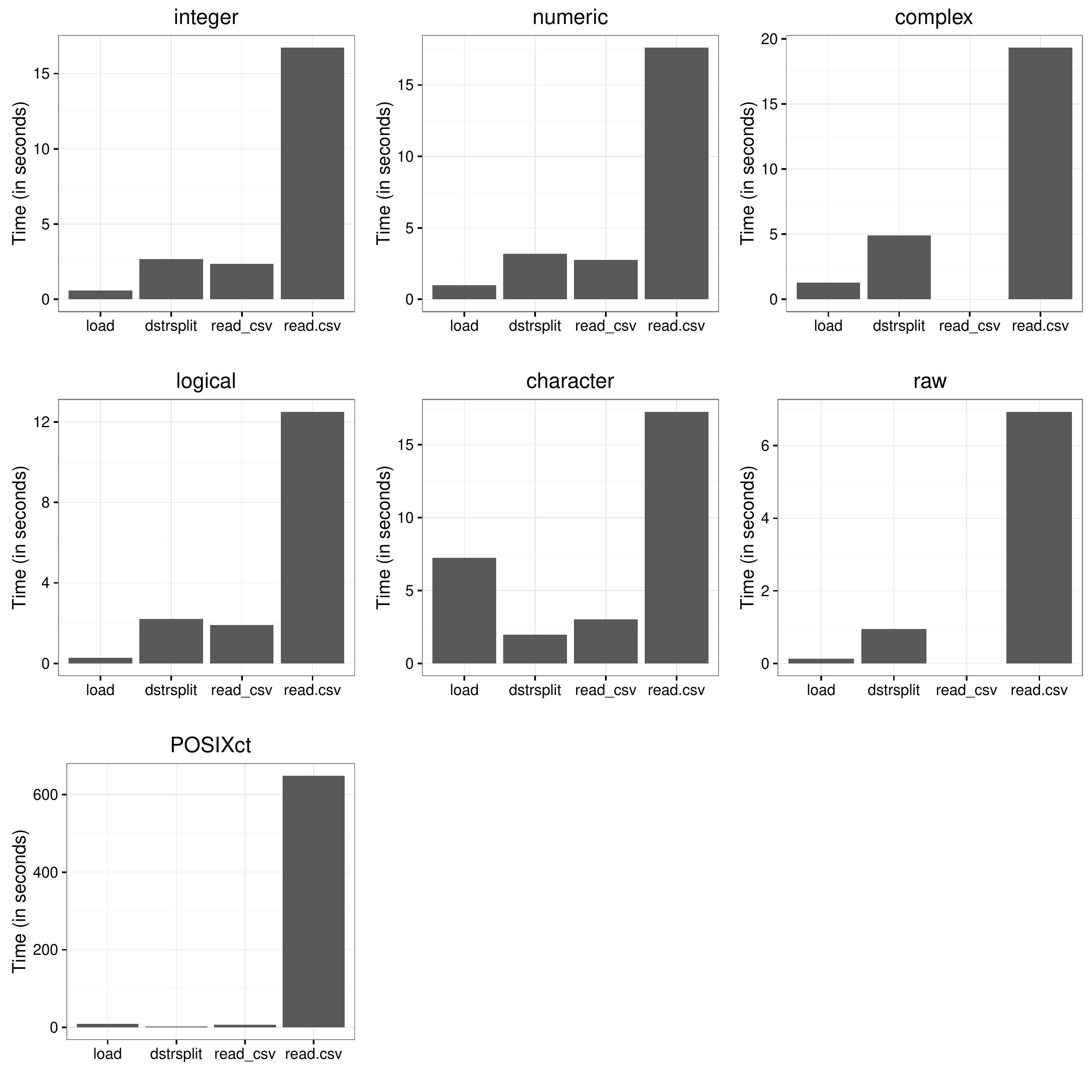}
\caption{Time to import \code{data.frame} by element type. Note \pkg{readr}
does not support \code{complex} and \code{raw} types.}
\label{fig:bench_data_frame}
\end{figure}

Figure \ref{fig:bench_data_frame} shows the time needed to import a 
data file with 1,000,000
rows and 25 columns using \code{load}, \code{dstrsplit}, \code{read\_csv} 
(from the \pkg{readr} package), and \code{read.table}. 
Imports were performed for each of \R's native types to see how their
different size requirements affect performance. The benchmarks show 
that, except for the POSIXct type, \code{load} is fastest. This
is unsurprising since \code{load} stores the binary representation 
of an \R~object and importing consists of copying the file to memory 
and registering the object in \R.

The \code{read\_csv}'s performance is very close to those of \code{iotools}.
When comparing with \code{read\_csv} we have found three things of note. First,
The difference in times are constant. As the number of
lines to read increases the slope of the read times are the same. Second,
where \code{read\_csv} maintains a slight edge in supported numerical types
on OS X, \code{iotools} has a slight edge on the Linux machines we tested.
Third, the \code{read\_csv} function was provided a connection explicitly in 
these benchmarks so all functions being examined are provided the same
input. When a file name is provided to \code{read\_csv}, it achieves
slightly better performance than the values shown here since it imports
from a memory-mapping of the file. 

\subsection{Processing and Checkpointing with \code{as.output}}

While optimized I/O operations are a convenience when performing
explorations and analyses on small data sets fitting in RAM, they
are an imperative when working with big data. In this class of data challenges 
we often deal with individual files whose aggregate is 
too large to fit in RAM. Furthermore, in distributed applications we may
need to load and process subsets from different machines, later combining
the results in some useful way. 

Let us assume we are tasked with finding the slope coefficents for
the linear regresssion
\begin{equation} \label{eqn:regression}
ArrDelay \sim DayOfWeek + DepTime + Month + DepDelay.
\end{equation}
The slope estimates
are formed by creating the model matrix, and applying the normal equations
to derive the coefficients. As a first task, we perform the simple 
preprocessing step of aggregating all of the files into a single file
holding the model matrix of the entire airline data set.
The slope coefficients will be calculated in a second, separate step. 

Separate processing and model fitting in this case are mostly for the 
sake of example. However, in many real-world data challenges it is 
a good idea. Separated steps provide {\em checkpointing} and if a problem 
arises while 
fitting the model, whether from a bug in the code or an interruption 
in computing services, the model matrix does not need to be 
recalculated. Also, the analysis can be changed based on the transformed
data, thereby saving a step for similar analyses. In the case
of our regression we can derive many different models involving the
described variables by including or excluding them in the model fitting step.

The example below shows how to write the model matrices to a single file,
named airline\_mm.io. However, we could have processed sets of files
just as easily with \pkg{iotools}. 
To emphasize \code{iotools} is complementary 
to existing packages, we will show its use with ``pipes'' 
included in the \pkg{tidyr} package \citep{tidyr}. The code
reads each of the airline files into a data frame
using \code{readAsRaw} and \code{dstrsplit}, normalizes the categorical
variables and transforms the departure times, creates a model matrix
from the resulting \code{data.frame}, strips the row names of the model
matrix, creates the text output representation, and writes it to
the output file.  The output connection is recycled in each iteration
of the loop thereby appending each year's data.

\begin{example}
> library(tidyr)
> 
> # The variables we'll use in the linear regression.
> form = ~ ArrDelay + DayOfWeek + DepTime + DepDelay + Month
> 
> # A function to normalize the categorical variables and turn 
> # departure time into minutes after midnight.
> normalize_df = function(x) {
+   names(x) = col_names
+   x$DayOfWeek = factor(x$DayOfWeek, levels=1:7)
+   x$Month = factor(x$Month, levels=1:12)
+   x$DepTime = sprintf("%04d", x$DepTime)
+   x$DepTime = as.numeric(substr(x$DepTime, 1, 2))*60 +
+     as.numeric(substr(x$DepTime, 3, 4))
+   x
+ }
> 
> # Remove the rownames from the output file.
> strip_rownames = function(x) {
+   rownames(x) = NULL
+   x
+ }
> 
> # Read the files and write a single model matrix.
> data_files = paste0(1988:2008, ".csv")
> out_file = file("airline_mm.io", "wb")
> for (data_file in data_files) {
+   data_file %>% readAsRaw %>%
+     dstrsplit(sep=",", skip=1, col_types=col_classes) %>% normalize_df %>%
+     model.matrix(form, .) %>% strip_rownames %>% as.output(sep=",") %>%
+     writeBin(out_file)
+ }
> close(out_file)
\end{example}

\subsection{Fitting the model with \code{mstrsplit} and \code{chunk.apply}}

With the model matrices created, The next step is to estimate 
the slope coefficients $\beta$ in the model 
\begin{equation} \label{eqn:linear_model}
Y = X \beta + \varepsilon,
\end{equation}
where $Y, \varepsilon \in \mathcal{R}^n$, and $\beta\in\mathcal{R}^d$, $n\ge d$;
each element of $\varepsilon$ is an i.i.d.
random variable with mean zero; and
$X$ is a matrix in $\mathcal{R}^{n \times d}$ with full column rank.
The analytic solution for estimating the slope coefficients, $\beta$,
is
\begin{equation} \label{eqn:lm_closed_form}
\widehat \beta = \left( X^T X \right)^{-1} X^T Y.
\end{equation}
Consider the row-wise partitioning (or chunking) of Equation 
\ref{eqn:linear_model}:
\begin{equation*}
\left[
  \begin{array}{c} Y_1 \\ Y_2 \\ \vdots \\ Y_r \end{array} \right] = 
  \left[ \begin{array}{c} X_1 \\ X_2 \\ \vdots \\ X_r
\end{array}
\right]
  \beta + \left[ \begin{array}{c} \varepsilon_1 \\ \varepsilon_2 \\ \vdots \\
                 \varepsilon_r \end{array} \right],
\end{equation*}
where
$Y_1, Y_2, ..., Y_r$, $X_1, X_2, ..., X_r$ and
$\varepsilon_1, \varepsilon_2, ...,  \varepsilon_r$ are
data partitions where each chunk is composed of subsets of
rows of the model matrix. Then Equation \ref{eqn:lm_closed_form}
can be expressed as 
\begin{equation}\label{blockwise_full}
\hat{\beta} = \left(\sum_{i=1}^r X_i^T X_i \right)^{-1}
\sum_{i=1}^r X_i^T Y_i.
\end{equation}
The matrices $X_i^TX_i$ and $X^T Y$ can be calculated on each chunk and then
summed to calculated the slope coefficients.
We remark computed solutions rarely use Equation \ref{eqn:lm_closed_form}
directly but rather use QR decompositions of $X$ for numerical
stability. In practice we have found the amount of numerical stability
gained does not warrant the QR calculation, especially when 
distinguishing nearly colinear variables is not critical.

Code to fit the model will need to read from airline\_mm.io in chunks
and where before data were read into a data frame, now we would like
to read data into a numeric matrix. Interestingly enough, this functionality
is not provided in base R or the \pkg{Matrix} \citep{matrix} package.
Traditionally, users who wanted to read matrices from disk either used the
\code{load}/\code{dget} function, forcing them to write using 
\code{save}/\code{dput}, or they could be read in as a data frame and 
then converted using the \code{as.matrix} function. The former approach
allows an \R~user to quickly import and export data but is not easily 
accessed from other computing environments. The latter requires a redundant
copy of the data. The \pkg{iotools} package fills this gap by providing
the \code{mstrsplit}, a matrix import function similar to \code{dstrsplit}

An implementation to fit a linear model out-of-core linear model shown below.
The \code{chunk.apply} function reads and processes chunks - in this 
case contiguous groups of rows in the model matrix.
The function takes as an argument a connection
or file, a function with a single parameter, and
a number of parallel processors to use. The function parameter
requires a single argument corresponding to 
the raw vector to be parsed by \code{dstrsplit} or \code{mstrsplit}.

\begin{example}
> # Get the factor expansion of the variables.
> mm_col_names = data_files[1] %>% read.csv.raw(header=TRUE, nrows=2) %>%
+   normalize_df %>% model.matrix(form, .) %>% colnames
> 
> ne_chunks = chunk.apply("airline_mm.io",
+   function(x) {
+     mm = mstrsplit(x, sep=",", type="numeric")
+     colnames(mm) = mm_col_names
+       list(xtx=crossprod(mm[,-2]),
+            xty=crossprod(mm[,-2], mm[,2, drop=FALSE]))
+   }, CH.MERGE=list, parallel=4)
> 
> xtx = Reduce(`+`, Map(function(x) x$xtx, ne_chunks))
> xty = Reduce(`+`, Map(function(x) x$xty, ne_chunks))
> 
> qr_xtx = qr(xtx)
> keep_vars = qr_xtx$pivot[1:qr_xtx$rank]
> 
> # The regression coefficients
> solve(xtx[keep_vars,keep_vars]) %*% xty[keep_vars]
                     [,1]
(Intercept)  0.5564085990
DayOfWeek2   0.5720431343
DayOfWeek3   0.8480978666
DayOfWeek4   1.2436976583
DayOfWeek5   1.0805744488
DayOfWeek6  -1.2235684080
DayOfWeek7  -0.9883340887
DepTime      0.0003022008
DepDelay     0.9329374752
Month2       0.2880436452
Month3      -0.2198123852
...
\end{example}

\begin{figure}[htbp]
\centering
\includegraphics[width=5in]{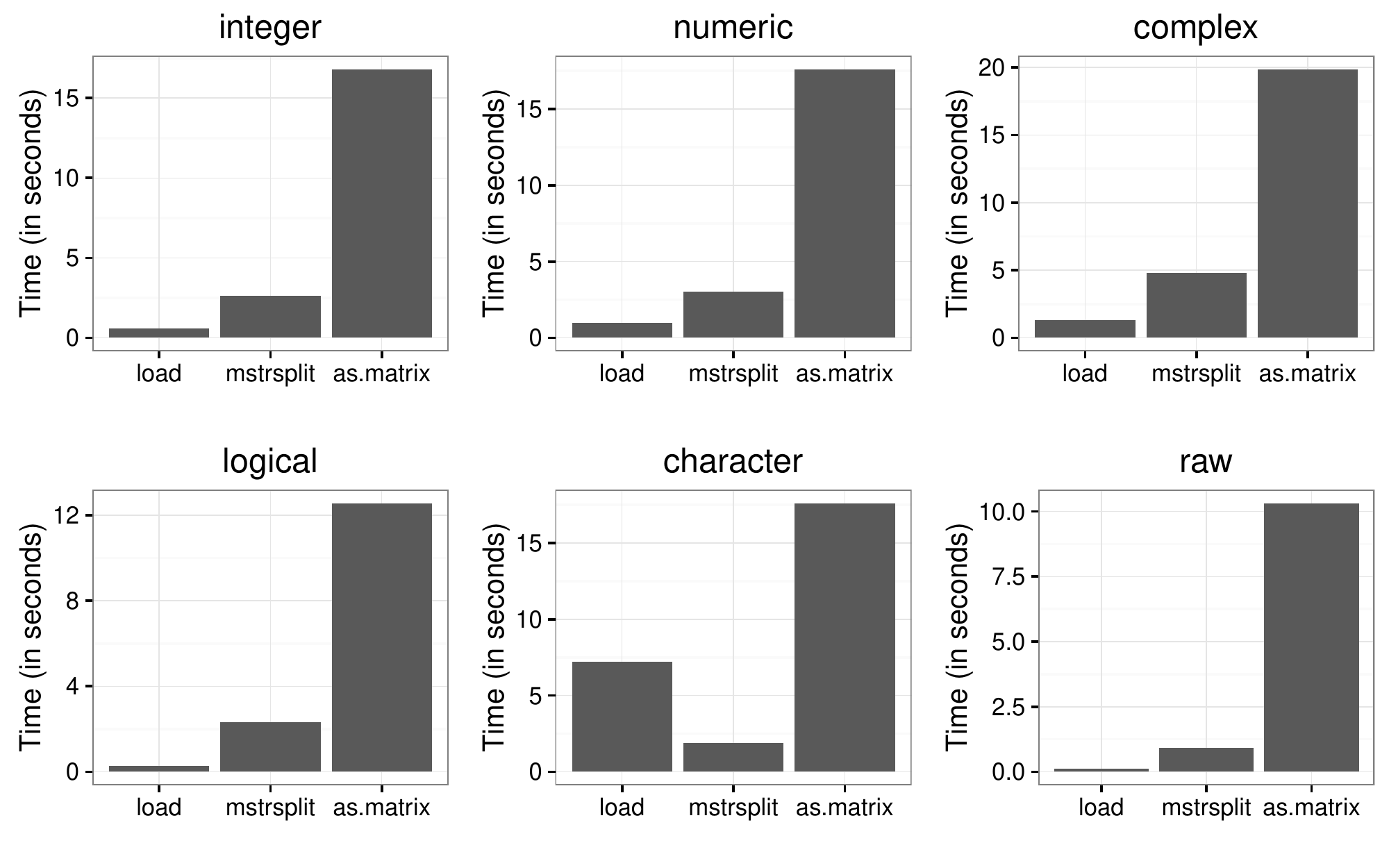}
\caption{Time to import a \code{matrix} by element type.}
\label{fig:bench_matrix}
\end{figure}

Figure \ref{fig:bench_matrix} compares the performance of \code{mstrsplit}
with \code{read.table} followed by a call to \code{as.matrix} along
binary importing using \code{load}. As with \code{dstrsplit}
\code{mstrsplit} outperforms the base \R's \code{read.table} benchmarks
by an order of magnitude and even outperforms \code{load} for 
\code{character} data.

\subsection{Parallel Processing of Chunks}

In the example above \code{xtx} and \code{xty} for each
chunk are calculated independently of any other chunk.  The \code{chunk.apply}
function includes a parameter, \code{parallel}, allowing the user to specify
the number of parallel processes, taking advantage of the embarrassingly
parallel nature of these calculations. However, it is worth noting 
parallelism in the \code{chunk.apply} function is slightly different than 
other functions such as \code{mclapply}.

Most parallel functions in \R~work by having worker processes receive 
data and an expression to compute. The master process 
initiates the computations and waits for them to complete.
For I/O-intensive computations this means either the master loads data
before initiating the computation or the worker processes load the
data. The former case is supported in \pkg{iotools} through 
iterator functions (\code{idstrsplit} and \code{imstrsplit}), which are 
compatible with the \pkg{foreach} package.
However, in this case, new tasks cannot be started until data has been
loaded for each of the workers. Loading the data on master process may become a 
bottleneck and it may require much more time to load the data than to process
it. The latter approach is also supported in \pkg{iotools} and 
ensures the master process is not a bottleneck
but if multiple worker processes on a single machine load a large amount
of data from the same disk then resource contention at the system
level may also cause excessive delays. The operating system has to service
multiple requests for data from the same disk having limited I/O 
capability.

A third option, implemented in \code{chunk.apply}, provides 
{\em pipeline parallelism}
where the master process sequentially loads data and then calls 
\code{mcparallel} to initiate the parallel computation. When the maximum
number of worker processes has been reached the master process 
{\em pre-fetches} the next chunk and then blocks on the result of
the running worker processes. When the result is returned a newly
created worker begins processing the pre-fetched data. In this way the 
master does not wait idly for worker processing and there is no 
resource contention since only the master is retrieving data.

Pipeline parallelism increases
execution throughput when the computation time is around the same order 
as the load time. When the overhead 
involved in initiating worker processes and getting their results
may overwhelm the computation time and parallel processing yield less
performant results.

\begin{figure}[htbp]
\centering
\includegraphics[width=5in]{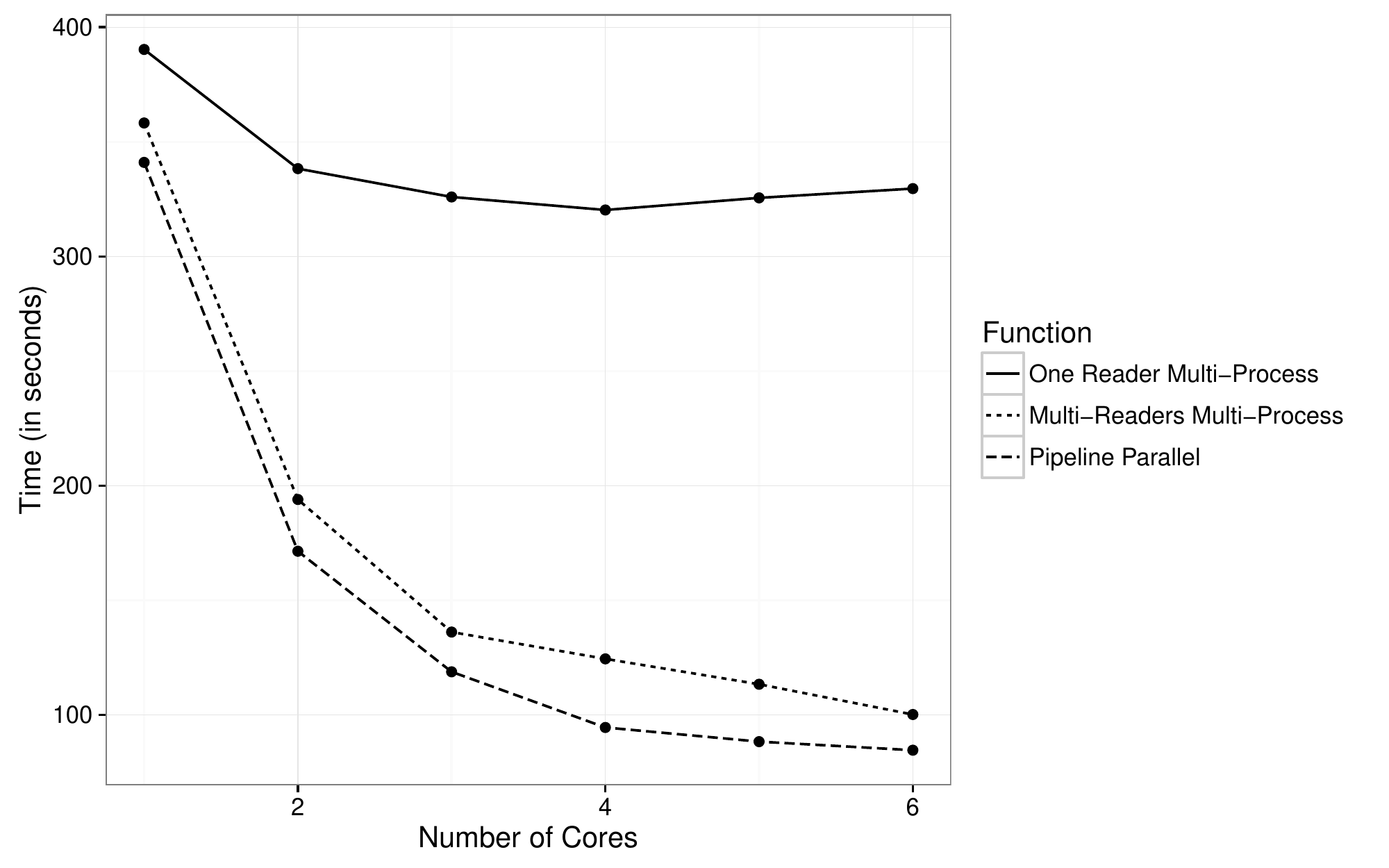}
\caption{Time to fit the linear model}
\label{fig:par_lm_times}
\end{figure}

Figure \ref{fig:par_lm_times} shows the times required to calculate 
$X^TX$ and $X^TY$ from the normal equations in the regression described
above using the three approaches described: all workers read, only 
the master reads, and pipeline parallel. Pipeline parallelism performs
best followed by all workers reading. 
It should be noted 
all workers reading 
will only be able to keep pace with pipeline parallelism as long as
there is sufficient hard-drive bandwidth and little contention from
multiple reads. As a result, the pipeline parallel approach is likely
a more general and therefore preferred strategy.

\section{Conclusion}

This paper presents the \pkg{iotools} package for the processing of 
data out-of-core and explores its use analyzing the Airline On-time data
set. The examples emphasize computing on a single machine. However it should
be noted \pkg{iotools} is by no means limited to this configuration.
The ``chunk'' functions are compatible with any object derived
from a \code{connection} and could therefore be used with compressed files
or even pipes and sockets. In fact, our current work uses \pkg{iotools}
as a building block for more tightly integrating \R~ in the Hadoop Streaming
and Spark frameworks. Early results show \pkg{iotools} achieves better
performance in processing terabyte- and even petabyte-scale data when 
compared to other existing packages.

\bibliography{arnold-kane-urbanek}

\def\noopsort#1{}
\begin{thebibliography}{17}
\providecommand{\natexlab}[1]{#1}
\providecommand{\url}[1]{\texttt{#1}}
\expandafter\ifx\csname urlstyle\endcsname\relax
  \providecommand{\doi}[1]{doi: #1}\else
  \providecommand{\doi}{doi: \begingroup \urlstyle{rm}\Url}\fi

\bibitem[Adler et~al.(2014)Adler, Gläser, Nenadic, Oehlschlägel, and
  Zucchini]{ff}
D.~Adler, C.~Gläser, O.~Nenadic, J.~Oehlschlägel, and W.~Zucchini.
\newblock \emph{ff: memory-efficient storage of large data on disk and fast
  access functions}, 2014.
\newblock URL \url{http://CRAN.R-project.org/package=ff}.
\newblock R package version 2.2-13.

\bibitem[Bates and Maechler(2014)]{matrix}
D.~Bates and M.~Maechler.
\newblock \emph{Matrix: Sparse and Dense Matrix Classes and Methods}, 2014.
\newblock URL \url{http://CRAN.R-project.org/package=Matrix}.
\newblock R package version 1.1-4.

\bibitem[Dean and Ghemawat(2008)]{Dean2008}
J.~Dean and S.~Ghemawat.
\newblock Map{R}educe: Simplified data processing on large clusters.
\newblock \emph{Commun. ACM}, 51\penalty0 (1):\penalty0 107--113, Jan. 2008.
\newblock ISSN 0001-0782.
\newblock \doi{10.1145/1327452.1327492}.
\newblock URL \url{http://doi.acm.org/10.1145/1327452.1327492}.

\bibitem[Guha et~al.(2012)Guha, Hafen, Rounds, Xia, Li, Xi, and
  Cleveland]{Guha2012}
S.~Guha, R.~Hafen, J.~Rounds, J.~Xia, J.~Li, B.~Xi, and W.~S. Cleveland.
\newblock Large complex data: divide and recombine (d{\&}r) with rhipe.
\newblock \emph{Stat}, 1\penalty0 (1):\penalty0 53--67, 2012.
\newblock ISSN 2049-1573.
\newblock \doi{10.1002/sta4.7}.
\newblock URL \url{http://dx.doi.org/10.1002/sta4.7}.

\bibitem[Hartigan(1975)]{Hartigan1975}
J.~A. Hartigan.
\newblock Necessary and sufficient conditions for asymptotic joint normality of
  a statistic and its subsample values.
\newblock \emph{The Annals of Statistics}, 3\penalty0 (3):\penalty0 573--580,
  05 1975.
\newblock \doi{10.1214/aos/1176343123}.
\newblock URL \url{http://dx.doi.org/10.1214/aos/1176343123}.

\bibitem[Kane et~al.(2013)Kane, Emerson, and Weston]{Kane2013}
M.~J. Kane, J.~Emerson, and S.~Weston.
\newblock Scalable strategies for computing with massive data.
\newblock \emph{Journal of Statistical Software}, 55\penalty0 (14):\penalty0
  1--19, 2013.
\newblock URL \url{http://www.jstatsoft.org/v55/i14/}.

\bibitem[Kleiner et~al.(2011)Kleiner, Talwalkar, Sarkar, and
  Jordan]{Kleiner2011}
A.~Kleiner, A.~Talwalkar, P.~Sarkar, and M.~I. Jordan.
\newblock A scalable bootstrap for massive data.
\newblock \emph{arXiv preprint arXiv:1112.5016}, 2011.

\bibitem[{Matloff}(2014)]{Matloff2014}
N.~{Matloff}.
\newblock {Software Alchemy: Turning Complex Statistical Computations into
  Embarrassingly-Parallel Ones}.
\newblock \emph{ArXiv e-prints}, Sept. 2014.

\bibitem[Mersmann(2014)]{microbenchmark}
O.~Mersmann.
\newblock \emph{microbenchmark: Accurate Timing Functions}, 2014.
\newblock URL \url{http://CRAN.R-project.org/package=microbenchmark}.
\newblock R package version 1.4-2.

\bibitem[{R Core Team}(2014)]{R}
{R Core Team}.
\newblock \emph{R: A Language and Environment for Statistical Computing}.
\newblock R Foundation for Statistical Computing, Vienna, Austria, 2014.
\newblock URL \url{http://www.R-project.org/}.

\bibitem[{Revolution Analytics}(2014)]{iterators}
{Revolution Analytics}.
\newblock \emph{iterators: Iterator construct for R}, 2014.
\newblock URL \url{http://CRAN.R-project.org/package=iterators}.
\newblock R package version 1.0.7.

\bibitem[RITA(2009)]{AirlineDataSet}
RITA.
\newblock The {A}irline on-time performance data set website, 2009.
\newblock URL \url{http://stat-computing.org/dataexpo/2009/}.
\newblock Research and Innovation Technology Administration, Bureau of
  Transportation Statistics.

\bibitem[{The Apache Software Foundation}(2013)]{Hadoop}
{The Apache Software Foundation}, 2013.
\newblock Apache Hadoop Streaming, available at \url{http://hadoop.apache.org}.

\bibitem[Weston and {Revolution Analytics}(2014)]{foreach}
S.~Weston and {Revolution Analytics}.
\newblock \emph{foreach: Foreach looping construct for R}, 2014.
\newblock URL \url{http://CRAN.R-project.org/package=foreach}.
\newblock R package version 1.4.2.

\bibitem[Wickham(2016)]{tidyr}
H.~Wickham.
\newblock \emph{tidyr: Easily Tidy Data with `spread()` and `gather()`
  Functions}, 2016.
\newblock URL \url{https://CRAN.R-project.org/package=tidyr}.
\newblock R package version 0.4.1.

\bibitem[Wickham and Francois(2015)]{readr}
H.~Wickham and R.~Francois.
\newblock \emph{readr: Read Tabular Data}, 2015.
\newblock URL \url{https://CRAN.R-project.org/package=readr}.
\newblock R package version 0.2.2.

\bibitem[Zaharia et~al.(2010)Zaharia, Chowdhury, Franklin, Shenker, and
  Stoica]{Spark}
M.~Zaharia, M.~Chowdhury, M.~J. Franklin, S.~Shenker, and I.~Stoica.
\newblock Spark: Cluster computing with working sets.
\newblock \emph{HotCloud}, 10:\penalty0 10--10, 2010.

\end{thebibliography}

\address{Taylor Arnold\\
  AT\&T Labs -- Statistics Research\\
  33 Thomas Street, NY \\
  USA}
\email{taylor@research.att.com}

\address{Michael J. Kane\\
  Yale University\\
  300 George Street, New Haven, CT\\
  USA}
\email{michael.kane@yale.edu}

\address{Simon Urbanek\\
  AT\&T Labs -- Statistics Research\\
  1 AT\&T Way Bedminster, NJ\\
  USA}
\email{urbanek@research.att.com}

\end{article}

\end{document}